\documentclass[11pt]{article}
\usepackage{amsmath,amsfonts,amsthm,fullpage,hyperref}

\newtheorem{theorem}{Theorem}[section]

\newtheorem{proposition}[theorem]{Proposition}
\newtheorem{lemma}[theorem]{Lemma}
\newtheorem{definition}[theorem]{Definition}

\title{Privacy and Mechanism Design}

\author{MALLESH M. PAI \\ Department of Economics, University of Pennsylvania \and AARON ROTH \\ Computer and Information Sciences, University of Pennsylvania}

\begin{document}


\maketitle

\begin{abstract}
This paper is a survey of recent work at the intersection of mechanism design and privacy. The connection is a natural one, but its study has been jump-started in recent years by the advent of \emph{differential privacy}, which provides a rigorous, quantitative way of reasoning about the costs that an agent might experience because of the loss of his privacy. Here, we survey several facets of this study, and differential privacy plays a role in more than one way. Of course, it provides us a basis for \emph{modeling} agent costs for privacy, which is essential if we are to attempt mechanism design in a setting in which agents have preferences for privacy.  It also provides a toolkit for controlling those costs. However, perhaps more surprisingly, it provides a powerful toolkit for controlling the stability of mechanisms in general, which yields a set of tools for designing novel mechanisms even in economic settings completely unrelated to privacy.
\end{abstract}

\section{Introduction}
Organizations such as census bureaus and hospitals have long maintained databases of personal information. However, with the advent of the Internet, many entities are now able to aggregate enormous quantities of personal and/or private information about individuals, with the intent to use it for financial gain or even malicious purposes. In reaction, several ``privacy advocacy'' groups have sprung up, with the intent to move US Congress and other lawmaking bodies to enact laws restricting the ability of private entities to collect and use personal information. Recent decisions by high-profile companies such as Facebook and Google have highlighted issues regarding privacy and brought them into public scrutiny.\footnote{Facebook has been accused of having a hard to use and frequently changing user interface for users privacy settings.}$^,$\footnote{Several Google projects, most recently their Glass project have drawn controversy, see, e.g. \url{http://blogs.wsj.com/digits/2013/05/16/congress-asks-google-about-glass-privacy/} .}

This interest in privacy is not solely or even largely motivated by the right to privacy as a basic desideratum. Increasingly, private information is explicitly being used for financial gain. In the recent past, companies have experimented with price discriminating against customers based on past purchase history,%
\footnote{See, for example, \url{http://www.cnn.com/2005/LAW/06/24/ramasastry.website.prices/}.}
technology choices,%
\footnote{See, for example, \url{http://www.cnn.com/2012/06/26/tech/web/orbitz-mac-users}.}
 or social profile.%
\footnote{For example, American Airlines offers customers a free day-pass to their premium lounges if they show they are influential on online social media via a ``Klout score.''. See \url{https://secure.fly.aa.com/klout/}.}
More broadly, there are concerns that the availability of such private information may influence important parts of an individual's life, e.g. access to health insurance or employment opportunities. As a result, issues related to privacy can have a large impact on individual welfare. An understanding of how agents' private data can be used in economic settings is therefore important to guiding policy.

Motivated by these issues, this article is part survey, part position paper and part progress report.  To formally study privacy, we have two ``toolboxes.'' The older literature is the large literature on information economics, game theory and mechanism design. The modern literature on ``differential privacy,'' on the other hand, gives a set of tools to reason about and control individual's costs for privacy loss. Combined, we can use these tools both to model settings in which agents have preferences toward privacy, and study mechanisms that trade off individual privacy with social goals. More surprisingly, the latter toolbox allows for the design of novel mechanisms in settings otherwise unrelated to privacy.

To briefly foreshadow the organization of this paper: in the next section, we quickly review the most basic aspects of differential privacy that we will use in this survey. We then study various recent contributions to mechanism design of two sorts. The first kind uses differential privacy as a tool to design novel mechanisms in settings where privacy is not a concern. The second considers the design of mechanisms in settings where agents have privacy concerns, i.e. the level of privacy the mechanism offers enters into agent's utility. Finally, we survey the (limited) literature that provides micro-foundations of preferences for privacy.

\section{Preliminaries}
This survey is chiefly (but not exclusively) interested in \emph{differential privacy} \cite{DMNS06}.
Let $\mathcal{T}$ denote some type space, and let $\mathcal{O}$ denote some outcome space. We will write $t \in \mathcal{T}^n$ to denote a vector of $n$ types, using the usual convention of indexing the $i$'th type by $t_i$, and the vector of all types \emph{excluding} the $i$'th type by $t_{-i}$. We will say that two type vectors $t, t' \in \mathcal{T}^n$ are \emph{neighbors} if there exists some index $i$ such that $t_{-i} = t'_{-i}$: in other words, $t$ and $t'$ only differ in their $i$'th index. We are now prepared to define differential privacy, which will be a property of \emph{randomized} mappings $M:\mathcal{T}^n\rightarrow \mathcal{O}$. We refer to these as \emph{mechanisms}.

\begin{definition}
A mechanism $M:\mathcal{T}^n\rightarrow \mathcal{O}$ is $\epsilon$-differentially private if for all pairs of neighboring type vectors $t, t' \in \mathcal{T}^n$, and for all functions $u:\mathcal{O}\rightarrow \mathbb{R}^+$:%
\footnote{We think of $\epsilon$ as being a small constant less than one, and so $\exp(\epsilon) \approx 1+\epsilon$.}
$$\mathbb{E}_{o \sim M(t)}[u(o)] \leq \exp(\epsilon) \mathbb{E}_{o \sim M(t')}[u(o)].$$
\end{definition}
Note that the `neighbor' relation is symmetric, so by definition, we also have the reverse inequality
$$\mathbb{E}_{o \sim M(t)}[u(o)] \geq \exp(-\epsilon) \mathbb{E}_{o \sim M(t')}[u(o)]$$

In other words, differential privacy promises that \emph{simultaneously}, for every possible utility function $u:\mathcal{O}\rightarrow \mathbb{R}^+$, the unilateral change of a single reported type $t_i$ to a mechanism can have only a small ($\approx 1+\epsilon$) multiplicative effect on the expected utility of the outcome drawn from the mechanism $M$.  We note that this definition is syntactically different from the standard definition of differential privacy \cite{DMNS06}, but is easily seen to be equivalent.

We will work with this version of the definition, which is particularly natural in the context of mechanism design. This version of the definition also makes it apparent why differential privacy corresponds to something that one would think of as ``privacy.'' It promises that \emph{regardless of your preferences}, your expected utility is not substantially changed if you decide to participate in the mechanism, compared to not participating (or, say, providing random data). Thus, given the choice to participate in a differentially private computation, you should be willing if given some (small) incentive to do so.%
\footnote{This incentive could take the form of a monetary payment, or could simply be the joy of furthering science, or the love of filling out forms.}

There is a large literature on differential privacy which we will not attempt to survey--- we direct the reader to \cite{DworkRoth} for an introduction to the area. Here, we mention just one differentially private mechanism: the exponential mechanism of \cite{MT07}.

\begin{definition}
The \emph{exponential mechanism} is defined by a range $\mathcal{R}$, a privacy parameter $\epsilon$, and a ``quality function'' $q:\mathcal{T}^n\times \mathcal{R}\rightarrow \mathbb{R}$ which has the property that for all pairs of neighboring type vectors $t, t' \in \mathcal{T}^n$, and for all $r \in \mathcal{R}: |q(t, r) - q(t', r)| \leq \Delta$. We refer to this constant $\Delta$ as the \emph{sensitivity} of $q$. Given an input $t \in \mathcal{T}^n$, the exponential mechanism outputs $r \in \mathcal{R}$ according to the distribution
$$r \propto \exp\left(\frac{\epsilon q(t, r)}{2\Delta}\right).$$
\end{definition}

The exponential mechanism is extremely useful due to the following theorem:
\begin{theorem}[\cite{MT07}]
\label{thm:exp}
The exponential mechanism is $\epsilon$-differentially private and with probability $1-\beta$ outputs some $r \in \mathcal{R}$ such that
$$q(t, r) \geq \max_{r^* \in \mathcal{R}}q(t,r^*) - \frac{2\Delta}{\epsilon}\left(\ln \frac{|\mathcal{R}|}{\beta}\right).$$
\end{theorem}
In other words, the exponential mechanism is a differentially private mechanism that outputs an element from the range that has quality score that is nearly as high as possible---excepting an additive term which is linear in the sensitivity of the quality score, and only logarithmic in the cardinality of the range of the mechanism.

\subsection{Differential Privacy as a Solution Concept}

Let us start by recalling a basic notion from mechanism design: dominant strategy truthfulness, also known as strategyproofness. Suppose that agents $i \in \{1,\ldots,n\}$ with types $t_i \in \mathcal{T}^n$ have utility functions $u_i:\mathcal{O}\rightarrow [0,1]$ over outcomes in $\mathcal{O}$ chosen by a mechanism $M$.
\begin{definition}
$M:\mathcal{T}^n\rightarrow \mathcal{O}$ is $\epsilon$-approximately dominant strategy truthful if for every player $i$, for every $t_{-i} \in \mathcal{T}^{n-1}$, and for every $t' \in \mathcal{T}$:
$$\mathbb{E}_{o \sim M(t_i,t_{-i})}[ u_i(o)] \geq \mathbb{E}_{o \sim M(t'_i,t_{-i})}[ u_i(o)] - \epsilon$$
%
\end{definition}

\cite{MT07} were the first to observe that differential privacy is a stronger guarantee than approximate truthfulness. Note that for $\epsilon \leq 1$, $\exp(\epsilon) \leq 1+2\epsilon$ and so the following proposition is immediate.
\begin{proposition}
If a mechanism $M$ is $\epsilon$-differentially private, then $M$ is also $2\epsilon$-approximately dominant strategy truthful.
\end{proposition}

As a solution concept, this has several robustness properties that strategy proof mechanisms do not. For example, the following is almost immediate from the definition of differential privacy: If $M_1$ and $M_2$ are both $\epsilon$-differentially private, and $f$ is any function (including the identity function), then $M_3$, defined as $M_3(t) = f(M_1(t),M_2(t))$ is $2\epsilon$-differentially private. This means in particular that the composition of two $\epsilon$-differentially private mechanisms remains $4\epsilon$-approximately dominant strategy truthful. In contrast, the incentive properties of general strategy proof mechanisms may not be preserved under composition.

Another useful property of differential privacy follows immediately from its definition: suppose that $t$ and $t' \in \mathcal{T}^n$ are not neighbors, but instead differ in $k$ indices. Then we have: $\mathbb{E}_{o \sim M(t)}[u(o)] \leq \exp(k\epsilon) \mathbb{E}_{o \sim M(t')}[u(o)]$. That is, changes in up to $k$ types changes the expected output by at most $\approx (1+k\epsilon)$, when $k \ll 1/\epsilon$. Therefore, differentially private mechanisms make truthful reporting a $2k\epsilon$-approximate dominant strategy \emph{even for coalitions of $k$ agents} -- i.e. differential privacy automatically provides robustness to collusion. Again, this is in contrast to general dominant-strategy truthful mechanisms, which in general offer no guarantees against collusion.

Notably, differential privacy allows for these properties in very general settings \emph{without the use of money!} In contrast, the set of exactly dominant strategy mechanisms when monetary transfers are not allowed is extremely limited.

We conclude with a drawback of using differential privacy as a solution concept as stated, first raised in \cite{NST12}: not only is truthfully reporting one's type an approximate dominant strategy, \emph{any report} is an approximate dominant strategy! That is, differential privacy makes the outcome approximately independent of any single agent's report. In some settings, this shortcoming can be alleviated. For example, suppose that $M$ is a differentially private mechanism, but that agent utility functions are defined to be functions both of the outcome of the mechanism, \emph{and} of the reported type of the agent: $u_i:\mathcal{O}\times\mathcal{T}\rightarrow [0,1]$. Suppose furthermore that for every outcome $o$, truthful reporting is a best response. In other words, for all $o$: $u_i(o,t) \geq \max_{t'_i \in \mathcal{T}}u_i(o, t')$. In this case, it is not hard to verify that the mechanism remains approximately dominant strategy truthful, but it is no longer the case that all reports are approximate dominant strategies.

\section{(Differential) Privacy as a Tool in Mechanism Design}
\label{sec:tool}

In this section, we show how the machinery of differential privacy can be used as a tool in designing novel mechanisms.

\subsection{Warmup: Digital Goods Auctions}
To warm up, let us consider a simple special case of the first application of differential privacy in mechanism design, the seminal \cite{MT07}.  Consider a \emph{digital goods auction}, i.e. one where the seller has an unlimited supply of a good with zero marginal cost to produce, for example a piece of software or other digital media. There are $n$ unit demand buyers for this good, each with unknown valuation $v_i \in [0,1]$. There is no prior on the bidder valuations, so a natural revenue benchmark is the revenue of the \emph{best fixed price}. At a price $p \in [0,1]$, each bidder $i$ with $v_i \geq p$ will buy. Therefore the total revenue of the auctioneer is
$$\textrm{Rev}(p,v) = p\cdot|\{i : v_i \geq p\}|.$$
The optimal revenue is the revenue of the best fixed price: $\textrm{OPT} = \max_p \textrm{Rev}(p,v)$. This setting is well studied--- \cite{BBHM05} give a dominant strategy truthful mechanism which achieves revenue at least $\textrm{OPT} - O(\sqrt{n})$.

We show how a simple application of the exponential mechanism achieves revenue at least $\textrm{OPT} - O\left(\tfrac{\log n}{\epsilon}\right)$. That is, the mechanism trades exact for approximate truthfulness, but achieves an exponentially better revenue guarantee. Of course, it also inherits the benefits of differential privacy discussed previously, such as resilience to collusion, and composability.

The idea is to select a price from the exponential mechanism, using as our ``quality score'' the revenue that this price would obtain. As we have defined it, the exponential mechanism is parameterized by some discrete range.%
\footnote{This is not necessary, but simplifies the exposition.}
Suppose we choose the range of the exponential mechanism to be $\mathcal{R} = \{\alpha, 2\alpha, \ldots, 1\}$? The size of the range is $|\mathcal{R}| = 1/\alpha$. What have we lost in potential revenue if we restrict ourselves to selecting a price from $\mathcal{R}$? It is not hard to see that
$$\textrm{OPT}_\mathcal{R} \equiv \max_{p \in \mathcal{R}}\textrm{Rev}(p,v) \geq \textrm{OPT} - \alpha n.$$
This is because if $p^*$ is the price that achieves the optimal revenue, and we use a price $p$ such that $p^*-\alpha \leq p \leq p^*$, every buyer who bought at the optimal price continues to buy, and provides us with at most $\alpha$ less revenue per buyer. Since there are at most $n$ buyers, the total lost revenue is at most $\alpha n$.

So how do we parameterize the exponential mechanism? We have a family of discrete ranges $\mathcal{R}$, parameterized by $\alpha$. For a vector of values $v$ and a price $p \in \mathcal{R}$, we define our quality function to be $q(v, p) = \textrm{Rev}(v,p)$. Observe that because each value $v_i \in [0,1]$, the \emph{sensitivity} of $q$ is $\Delta = 1$: changing one bidder valuation can only change the revenue at a fixed price by at most $v_i \leq 1$. Therefore, if we require $\epsilon$-differential privacy, from Theorem \ref{thm:exp}, we get that with high probability, the exponential mechanism returns some price $p$ such that
$$\textrm{Rev}(p,v) \geq \left(\textrm{OPT} -\alpha n\right) - O\left(\frac{1}{\epsilon} \ln\left(\frac{1}{\alpha}\right)\right).$$
Choosing our discretization parameter $\alpha$ to minimize the two sources of error, we find that this mechanism with high probability finds us a price that achieves revenue
$$\textrm{Rev}(p,v) \geq \textrm{OPT} - O\left(\frac{\log n}{\epsilon}\right).$$

Note that if we take (e.g.) $\epsilon = 1/\log(n)$, then we obtain a mechanism that is asymptotically exactly truthful (i.e. as the market grows large, the approximation to truthfulness becomes exact), while still achieving revenue at least $(1-o(1))\textrm{OPT}$, so long as $\textrm{OPT}$ grows more quickly than $\log(n)^2$ with the size of the population $n$.

Finally, notice that we could make the reported value $v_i$ of each agent $i$ binding. In other words,  we could allocate an item to agent $i$ and extract payment of the selected posted price $p$ whenever $v_i \geq p$. If we do this, the mechanism is approximately truthful, because the price is picked using a differentially private mechanism. Additionally, it is not the case that \emph{every} report is an approximate dominant strategy: if an agent over-reports, she may be forced to buy the good at a price higher than her true value.

\subsection{Approximately Truthful Equilibrium Selection Mechanisms}
We now consider the problem of approximately truthful equilibrium selection, studied in \cite{KPRU12}. Roughly speaking, the problem is as follows: suppose we are given a game in which each player knows their own payoffs, but not others' payoffs. The players therefore do not know the equilibrium structure of this game. Even if they did, there might be multiple equilibria, with different agents preferring different equilibria. Can a mechanism offered by an intermediary incentivize agents to truthfully report their utilities and follow the equilibrium it selects?

For example, imagine a city in which (say) Google Navigation is the dominant service. Every morning, each person enters their starting point and destination, receives a set of directions, and chooses his/ her route according to those directions. Is it possible to design a navigation service such that: (1) Each agent is incentivized to report truthfully, and (2) then follow the driving directions provided? Both misreporting start and end points, and truthfully reporting start and end points, but then following a different (shorter) path are to be disincentivized.

Intuitively, our two desiderata are in conflict. In the commuting example above, if we are to guarantee that every player is incentivized to truthfully follow their suggested route, then we must compute an equilibrium of the game in question given players' reports. On the other hand, to do so, our suggested route to some player $i$ must depend on the reported location/ destination pairs of other players. This tension will pose a problem in terms of incentives: if we compute an equilibrium of the game given the reports of the players, an agent can potentially benefit by misreport, causing us to compute an equilibrium of the wrong game.

This problem would be largely alleviated, however, if the report of agent $i$ only has a tiny affect on the actions of agents $j \neq i$. In this case, agent $i$ could hardly gain an advantage through his effect on other players. Then, assuming that everyone truthfully reported their type, the mechanism would compute an equilibrium of the correct game, and by definition, each agent $i$ could do no better than follow the suggested equilibrium action. In other words, if we could compute an approximate equilibrium of the game under the constraint of \emph{differential privacy}, then truthful reporting, followed by taking the suggested action of the coordination device would be a Nash equilibrium. A moment's reflection reveals that the goal of privately computing an equilibrium is not possible in small games, in which an agent's utility is a highly sensitive function of the actions (and hence utility functions) of the other agents.%
\footnote{Positive results are not beyond hope in small games for slightly different settings. See, e.g. \cite{dziuda2012coordination}.}
But what about in large games?

Formally, suppose we have an $n$ player game with action set $\mathcal{A}$, and each agent with type $t_i$ has a utility function $u_i:\mathcal{A}^n\rightarrow [0,1]$. We say that this game is $\Delta$-large if for all players $i \neq j$, vectors of actions $a \in \mathcal{A}^n$, and pairs of actions $a_j, a_j' \in \mathcal{A}$:
$$\left|u_i(a_j, a_{-j}) - u_i(a'_j, a_{-j})\right| \leq \Delta.$$
In other words, if some agent $j$ unilaterally changes his action, then his affect on the payoff of any other agent $i \neq j$ is at most $\Delta$. Note that if agent $j$ changes his own action, then his payoff can change arbitrarily. Many games are ``large'' in this sense. In the commuting example above, if Alice changes her route to work she may substantially increase or decrease her commute time, but will only have a minimal impact on the commute time of any other agent Bob. The results in this section are strongest for $\Delta = O(1/n)$, but hold more generally.

First we might ask whether we need privacy at all--- could it be the case that in a large game, any algorithm which computes an equilibrium of a game defined by reported types has the stability property that we want? The answer is no. As a simple example, consider $n$ people who must each choose whether to go to the beach (B) or the mountains (M). People privately know their types--- each person's utility depends on his own type, his action, and the fraction of other people $p$ who go to the beach. A Beach type gets a payoff of $10p$ if he visits the beach, and $5(1-p)$ if he visits the mountain. A mountain type gets a payoff $5p$ from visiting the beach, and $10(1-p)$ from visiting the mountain. Note that this is a large game--- each player's payoffs are insensitive in the actions of others.  Further, note that ``everyone visits beach'' and ``everyone visits mountain'' are both equilibria of the game, regardless of the realization of types. Consider the mechanism that attempts to implement the following social choice rule--- ``if the number of beach types is less than half the population, send everyone to the beach, and vice versa.'' It should be clear that if mountain types are just in the majority, then each mountain type has an incentive to misreport as a beach type; and vice versa. As a result, even though the game is ``large'' and agents' actions do not affect others' payoffs significantly, simply computing equilibria from reported type profiles does not in general lead to even approximately truthful mechanisms.

Nevertheless, \cite{KPRU12} are able to give a mechanism with the following property: it elicits the type $t_i$ of each agent, and then computes an $\alpha$-approximate correlated equilibrium of the game defined by the reported types.%
\footnote{A correlated equilibrium is defined by a joint distribution on profiles of actions, $\mathcal{A}^n$. For an action profile $a$ drawn from the distribution, if agent $i$ is told only $a_i$, then playing action $a_i$ is a best response given the induced conditional distribution over $a_{-i}$. An $\alpha$-approximate correlated equilibrium is one where deviating improves an agent utility by at most $\alpha$.}
It draws an action profile $a \in \mathcal{A}^n$ from the correlated equilibrium, and reports action $a_i$ to each agent $i$. The algorithm has the guarantee that simultaneously for all players $i$, the joint distribution $a_{-i}$ on reports to all players \emph{other than $i$} is differentially private in the reported type of agent $i$. This guarantee is sufficient for approximate truthfulness, because it means that agent $i$ cannot substantially change the distribution on actions induced on \emph{the other players} by misreporting his own type.

More specifically, when the mechanism of \cite{KPRU12} computes an $\alpha$-approximate correlated equilibrium while satisfying $\epsilon$-differential privacy, every agent following the honest behavior (i.e. first reporting their true type, then following their suggested action) forms an $(2\epsilon+\alpha)$-approximate Nash equilibrium. This is because, by privacy, reporting your true type is a $2\epsilon$-approximate dominant strategy, and given that everybody reports their true type, the mechanism computes an $\alpha$-approximate correlated equilibrium of the true game, and hence by definition, following the suggested action is an $\alpha$-approximate best response. \cite{KPRU12} give mechanisms for computing $\alpha$-approximate equilibrium in large games with $\alpha = O\left(\frac{1}{\sqrt{n}\epsilon}\right)$. Therefore, by setting $\epsilon = O\left(\frac{1}{n^{1/4}}\right)$, this gives an $\eta$-approximately truthful equilibrium selection mechanism for
$$\eta = 2\epsilon+\alpha = O\left(\frac{1}{n^{1/4}}\right).$$
In other words, it gives a mechanism for coordinating equilibrium behavior in large games that is asymptotically truthful in the size of the game, all without the need for monetary transfers.

 \subsection{Obtaining Exact Truthfulness}
 \label{sec:restrict}
 So far we have discussed mechanisms that are \emph{asymptotically truthful} in large population games. However, what if we want to insist on mechanisms that are \emph{exactly} dominant strategy truthful, while maintaining some of the nice properties enjoyed by our mechanisms so far: for example, that the mechanisms do not need to be able to extract monetary payments? Can differential privacy help here? It can---in this section, we discuss a special case of a framework laid out by \cite{NST12} which uses differentially private mechanisms as a building block towards designing exactly truthful mechanisms without money.

 The basic idea is simple and elegant. As we have seen, the exponential mechanism can often give excellent utility guarantees while preserving differential privacy. This doesn't yield an exactly truthful mechanism, but it gives every agent very little incentive to deviate from truthful behavior. What if we could pair this with a second mechanism which need not have good utility guarantees, but gives each agent a strict positive incentive to report truthfully, i.e. a mechanism that essentially only punishes non-truthful behavior? Then, we could randomize between running the two mechanisms. If we put enough weight on the punishing mechanism, then we inherit its strict-truthfulness properties. The remaining weight that is put on the exponential mechanism contributes to the utility properties of the final mechanism. The hope is that since the exponential mechanism is approximately strategy proof to begin with, the randomized mechanism can put small weight on the strictly truthful punishing mechanism, and therefore will have good utility properties.

 To design punishing mechanisms, \cite{NST12} work in a slightly non-standard environment. Rather than simply picking an outcome, they model a mechanism as picking an outcome, and then an agent as choosing a \emph{reaction} to that outcome, which together define his utility. They then give the mechanism the power to \emph{restrict the reactions allowed by the agent based on his reported type}. Formally, they work in the following framework:

 \begin{definition}[The Environment \cite{NST12}]
An environment is a set $N$ of $n$ players, a set of types $t_i \in \mathcal{T}$, a finite set $\mathcal{O}$ of outcomes, a set of reactions $R$ and a utility function $u_i:T\times \mathcal{O} \times R \rightarrow [0,1]$ for each agent $i$.
\end{definition}

We write  $r_i(t,s, \hat{R}_i) \in \arg\max_{r \in \hat{R}_i}u_i(t, s, r)$ to denote $i$'s optimal reaction among choices $\hat{R}_i \subseteq R$ to alternative $s$ if he is of type $t$.

A direct revelation mechanism $\mathcal{M}$ defines a game which is played as follows:
\begin{enumerate}
\item Each player $i$ reports a type $t'_i \in \mathcal{T}$.
\item The mechanism chooses an alternative $s \in \mathcal{O}$ and a subset of reactions for each player $\hat{R}_i \subseteq R$.
\item Each player chooses a reaction $r_i \in \hat{R}_i$ and experiences utility $u_i(t_i, s, r_i)$.
\end{enumerate}
Agents play so as to maximize their own utility. Note that since there is no further interaction after the 3rd step, rational agents will pick $r_i = r_i(t_i, s, \hat{R}_i)$, and so we can ignore this as a strategic step. Let $\mathcal{R} = 2^{R}$. Then a mechanism is a randomized mapping $\mathcal{M}:\mathcal{T}\rightarrow \mathcal{O}\times \mathcal{R}^n$.

Let us consider the utilitarian welfare criterion: $F(t, s, r) = \frac{1}{n}\sum_{i=1}^n u_i(t_i, s, r_i)$, Note that this has sensitivity $\Delta = 1/n$, since each agent's utility lies in the range $[0,1]$. Hence, if we simply choose an outcome $s$ and allow each agent to play their best response reaction, the exponential mechanism is an $\epsilon$-differentially private mechanism, which by Theorem \ref{thm:exp},  achieves social welfare at least $\textrm{OPT} - O\left(\frac{\log |\mathcal{O}|}{\epsilon n}\right)$ with high probability. Let us denote this instantiation of the exponential mechanism, with quality score $F$, range $\mathcal{O}$ and privacy parameter $\epsilon$, as $\mathcal{M}_\epsilon$.

The idea is to randomize between the exponential mechanism (with good social welfare properties) and a strictly truthful mechanism which punishes false reporting (but with poor social welfare properties). If we mix appropriately, then we will get an exactly truthful mechanism with reasonable social welfare guarantees.

Here is one such punishing mechanism which is simple, but not necessarily the best for a given problem:
\begin{definition}[\cite{NST12}]
The commitment mechanism $M^P(t')$ selects $s\in \mathcal{O}$ uniformly at random and sets $\hat{R}_i = \{r_i(t_i', s, R_i)\}$, i.e. it picks a random outcome and forces everyone to react as if their reported type was their true type. \end{definition}
Define the \emph{gap} of an environment as
$$\gamma = \min_{i, t_i \neq t_i',t_{-i}}\max_{s \in \mathcal{O}}\left(u_i(t_i, s, r_i(t_i, s, R_i)) - u_i(t_i, s, r_i(t_i', s, R_i))\right),$$
i.e. $\gamma$ is a lower bound over players and types of the worst-case cost (over $s$) of mis-reporting. Note that for each player, this worst-case is realized with probability at least $1/|\mathcal{O}|$. Therefore we have the following simple observation:
\begin{lemma}
For all $i$, $t_i, t_i', t_{-i}$:
$$u_i(t_i, \mathcal{M}^P(t_i, t_{-i})) \geq  u_i(t_i, \mathcal{M}^P(t_i', t_{-i})) + \frac{\gamma}{|\mathcal{O}|}$$
\end{lemma}
Note that the commitment mechanism is strictly truthful: every individual has at least a $\frac{\gamma}{|\mathcal{O}|}$ incentive not to lie.

This suggests an exactly truthful mechanism with good social welfare guarantees:
\begin{definition}
The punishing exponential mechanism $\mathcal{M}_\epsilon^P(t)$ defined with parameter $0 \leq q \leq 1$ is, selects the exponential mechanism $\mathcal{M}_\epsilon(t)$ with probability $1-q$ and the punishing mechanism $\mathcal{M}^P(t)$ with complementary probability $q$.
\end{definition}

The following two theorems from \cite {NST12} show incentive and social welfare properties of this mechanism.
\begin{theorem}
If $2\epsilon \leq \frac{q\gamma}{|\mathcal{O}|}$ then $\mathcal{M}^P_\epsilon$ is strictly truthful.
\end{theorem}


\begin{theorem}
\label{thm:utility}
For sufficiently large $n$, $M_\epsilon^P$ achieves social welfare at least
$$\textrm{OPT} - O\left(\sqrt{\frac{|\mathcal{O}|\log |\mathcal{O}|}{\gamma n}}\right)$$
\end{theorem}

Note that this mechanism is truthful without the need for payments!

Let us now consider an application of this framework: the facility location game. Suppose that a city wants to build $k$ hospitals to minimize the average distance between each citizen and their closest hospital. To simplify matters, we make the mild assumption that the city is built on a discretization of the unit line.\footnote{If this is not the case, we can easily raze and then re-build the city.} Formally, for all $i$ let $L(m) = \{0, \frac{1}{m}, \frac{2}{m}, \ldots, 1\}$
denote the discrete unit line with step-size $1/m$. $|L(m)| = m+1$. Let $\mathcal{T} = R_i = L(m)$ for all $i$ and let $|\mathcal{O}| = L(m)^k$. Define the utility of agent $i$ to be:
$$u_i(t_i, s, r_i) = \left\{
                       \begin{array}{ll}
                         -|t_i-r_i|, & \hbox{If $r_i \in s$;} \\
                         -1, & \hbox{otherwise.}
                       \end{array}
                     \right.$$
In other words, agents are associated with points on the line, and an outcome is an assignment of a location on the line to each of the $k$ facilities. Agents can react to a set of facilities by deciding which one to go to, and their cost for such a decision is the distance between their own location (i.e. their type) and the facility that they have chosen.
Note that $r_i(t_i, s)$ is here the closest facility $r_i \in s$.

We can instantiate Theorem \ref{thm:utility}. In this case, we have: $|\mathcal{O}| = (m+1)^k$ and $\gamma = 1/m$, because any two positions $t_i\neq t_i'$ differ by at least $1/m$. Hence, we have:
\begin{theorem}[\cite{NST12}]
$M_\epsilon^P$ instantiated for the facility location game is strictly truthful and achieves social welfare at least:
$$\textrm{OPT} -   O\left(\sqrt{\frac{km(m+1)^k\log m}{n}}\right)$$
\end{theorem}
This is already very good for small numbers of facilities $k$, since we expect that $\textrm{OPT} = \Omega(1)$. We note that for the facility location problem, \cite{NST12} derive a superior bound using a more refined argument.

\section{The Value of Privacy}
\label{sec:value}

In the previous section, we saw that differential privacy can be useful as a tool to design mechanisms, \emph{for agents who care only about the outcome chosen by the mechanism}. We here primarily viewed privacy as a tool to accomplish goals in traditional mechanism design.  As a side affect, these mechanisms also preserved the privacy of the reported player types.  Is this itself a worthy goal? \emph{Why} might we want our mechanisms to preserve the privacy of agent types?

A bit of reflection reveals that agents might care about privacy. Indeed, basic introspection suggests that in the real world, agents value the ability to keep certain ``sensitive'' information private, for example, health information or sexual preferences. In this section, we consider the question of how to model this value for privacy, and various approaches taken in the literature.

A first option is to just model value for privacy as a part of the agent's preferences. At one level, this is a satisfactory approach. Agents do seem to value privacy, and this is in the spirit of economic modeling ``\textit{De Gustibus non disputandum est}.'' However, such a ``reduced form'' approach may not be helpful in policy analysis.

Recall our original motivation for differential privacy--- it quantifies the worst case harm that can befall an agent from revealing his private data. A structural model of how the agent evaluates this harm may, therefore, be helpful in understanding both the individual value of privacy and the social value of privacy policies. For example, consider how an agent values the privacy of his health information. Consider two scenarios, one where health insurers or potential employers can discriminate based on an agent's health history and another where they cannot  do so.%
\footnote{As an aside, we should point out that simply banning discrimination on a certain attribute may not be sufficient to prevent discrimination using other information correlated with that attribute. See for example \cite{chan2003does}, and a related definition of fairness \cite{DHPRZ12}}
\textit{Ceteris paribus}, it seems reasonable that the dis-utility he suffers from his health information being made public is different in these two scenarios. A more structural model of preferences for privacy may therefore be more appropriate for understanding, e.g., the social value of privacy policies.

There have been a few notable papers that study privacy policy in dynamic models.%
\footnote{The broader field of information economics studies the value of information a variety of settings too vast to survey here. We restrict attention to papers the explicitly study the value of privacy policies.}
Agents' preferences for privacy in a period derive from how other players can use the information revealed against the agent in future periods.

Most of the papers we survey are motivated by repeat purchasers in electronic commerce settings. Information about purchases made by an agent in a setting with limited privacy can be used to learn about his `payoff type,' and therefore better price discriminate subsequently. An agent understands this and may distort his early purchases, depending on the privacy policy. Similarly, pricing by a profit maximizing seller also depends on the privacy policy. The trade-offs between various privacy policies can be studied by computing the equilibrium welfare and revenue under these policies.

An early paper in this area is \cite{taylor2004consumer}. He studies a setting where consumers purchase from firm $1$ in period $1$ and then firm $2$ in period $2$. Consumers have additive preferences for the two goods and may have either a high or low value for each good. These values are privately known to them, and for any given customer the two are unconditionally correlated. The paper studies three scenarios. In the first, firm $1$ must keep purchases by the consumers private. In the second, firm $1$ can sell this information to firm $2$, but consumers are unsophisticated. In other words consumers' purchase decisions in period $1$ are myopic, not taking into account how this will influence the prices they are offered in period $2$. The author shows that firms fare well in the latter relative to the former. Finally, the author considers the equilibrium of a model where consumers are strategic rather than unsophisticated. In this setting, consumers may strategically reduce demand in the first period. This undermines the market for consumer information, and the firms may prefer to commit to a policy of no market for consumer information.

Another early seminal paper is that of \cite{calzolari2006optimality}. They study a more general setting where an agent with private information sequentially contracts with two principals. The upstream principal may sell information to the downstream principal after contracting with the agent. This `privacy policy' is a part of the contract offered by the upstream principal, and he can commit to this. The agent is sophisticated and takes this into account. They provide a general characterization of settings in which the upstream principal offers full privacy. This hinges on three conditions being satisfied. Firstly, they require that the agent's trade with the downstream principal is not directly payoff relevant to the upstream principal. Secondly, they require that the agent's valuations across the two principals be positively correlated. Finally, they require that the preferences in the downstream relationship are separable so that the do not depend on the upstream level of trade. If any of these conditions are violated, full privacy need not remain optimal. Surprisingly, the paper shows that the agent may strictly prefer disclosure. In other words, the equilibrium value of privacy may be negative, which runs counter to our intuitions about the value of privacy.

\cite{conitzer2012hide} study a setting where a monopolist seller in a two-period model cannot commit to future prices.%
\footnote{See also \cite{acquisti2005conditioning} for a related study.}
Each agent has unit demand in each period, and a private value. In the second period, therefore, the monopolist conditions the price he offers on whether or not the agent bought in the first period. Intuitively, an agent who bought in the first period will face a higher price. The agent realizes this and therefore may not buy in the first period even if it is myopically optimal. In the model, agents may be able to `buy' privacy, i.e. avoid being identified as past customers, but possibly at a cost. On a similar vein to the previous two papers, they note that in this setting, if this privacy is available for free, all consumers will choose to purchase it, but in equilibrium this will be worse for the agent and better for the monopolist. Increasing the cost of anonymity can actually benefit consumers.

Finally, in a recent paper, \cite{bergemann2013limits} study price discrimination by a monopolist who has some exogenously specified information additional to the prior distribution of the buyers' type. The change in consumer and producer surplus from this additional information can thus be thought of as the value of privacy of this information.

This suggests that there is much to be done in terms of modeling and understanding preferences for privacy. Our basic intuition, i.e., that private information can be used to price discriminate against an agent, and therefore privacy is ``good'' for an agent, is not reflected in models where information is revealed by strategic purchases. Indeed, these papers unambiguously suggest that with strategic agents, the value of privacy is negative!

\section{Mechanism Design for Privacy Aware Agents}
\label{sec:aware}
Having established that agents might have preferences for privacy, it is worth considering the design of mechanisms that preserve privacy \emph{as an additional goal}, even for tasks such as, e.g. welfare maximization that we can already solve non-privately. As we will see, it is indeed possible to generalize the VCG mechanism to \emph{privately} approximately optimize social welfare in \emph{any} social choice problem, with a smooth trade-off between the privacy parameter and the approximation parameter, all while guaranteeing exact dominant strategy truthfulness.

However, we might wish to go further. In the presence of agents with preferences for privacy, if we wish to design truthful mechanisms, we must somehow model their preferences for privacy in their utility function, and then design mechanisms which are truthful with respect to these new ``privacy aware'' utility functions. As we have seen with differential privacy, it is most natural to model privacy as a property of the mechanism itself. Thus, our utility functions are not merely functions of the outcome, but functions of the outcome and of the mechanism itself. In almost all models, agent utilities for outcomes are treated as linearly separable, that is, we will have for each agent $i$,
$$u_i(o, \mathcal{M}, t) \equiv \mu_i(o) - c_i(o, \mathcal{M}, t).$$
Here $\mu_i(o)$ represents agent $i$'s utility for outcome $o$ and $c_i(o, \mathcal{M},t)$ the (privacy) cost that agent $i$ experiences when outcome $o$ is chosen with mechanism $\mathcal{M}$.

We will first consider perhaps the simplest (and most na\"ive) model for the privacy cost function $c_i$, following \cite{GR11}. Recall that for $\epsilon \ll 1$, differential privacy promises that for each agent $i$, and for every possible utility function $f_i$, type vector $t \in \mathcal{T}^n$, and deviation $t' \in \mathcal{T}$:
$$|\mathbb{E}_{o \sim M(t_i,t_{-i})}[ f_i(o)] - \mathbb{E}_{o \sim M(t'_i,t_{-i})}[ f_i(o)]| \leq \approx \epsilon \mathbb{E}_{o \sim M(t)}[ f_i(o)].$$
If we view $f_i$ as representing the ``expected future utility'' for agent $i$, it is therefore natural to model agent $i$'s cost for having his data used in an $\epsilon$-differentially private computation as being linear in $\epsilon$. That is, we think of agent $i$ as being parameterized by some value $v_i \in \mathbb{R}$, and take:
$$c_i(o, \mathcal{M}, t) = \epsilon v_i$$
where $\epsilon$ is the smallest value such that $\mathcal{M}$ is $\epsilon$-differentially private. Here we imagine $v_i$ to represent a quantity like $\mathbb{E}_{o \sim M(t)}[ f_i(o)]$. In this setting, $c_i$ does not depend on the outcome $o$ or the type profile $t$.

Using this na\"ive privacy measure, we discuss a basic problem in private data analysis: how to collect the data, when the owners of the data value their privacy and insist on being compensated for it. In this setting, there is no ``outcome'' that agents value, other than payments, there is only dis-utility for privacy loss. We will then discuss shortcomings of this (and other) measures of the dis-utility for privacy loss, as well as privacy in more general mechanism design settings when agents \emph{do} have utility for the outcome of the mechanism.

Our discussion here is in the context of a specific setting, i.e., the sensitive surveyor's problem. \cite{gradwohl2012privacy} considers the abstract problem of what social choice functions can be implemented when agents have preferences for privacy. He shows that extensive game forms are useful when agents have privacy concerns. We do not discuss this paper here, and refer interested readers to the original manuscript for details.

\subsection{A Private Generalization of the VCG Mechanism.}
Suppose we have a general social choice problem, defined by an outcome space $\mathcal{O}$, and a set of agents $N$ with arbitrary preferences over the outcomes given by $u_i:\mathcal{O}\rightarrow [0,1]$. We might want to choose an outcome $o \in \mathcal{O}$ to maximize the \emph{social welfare} $F(o) = \frac{1}{n}\sum_{i=1}^n u_i(o)$. It is well known that in any such setting, the \emph{VCG} mechanism can implement the outcome $o^*$ which exactly maximizes the social welfare, while charging payments that make truth-telling a dominant strategy. What if we want to achieve the same result, while also preserving privacy? How must the privacy parameter $\epsilon$ trade off with our approximation to the optimal social welfare?

Recall that we could use the exponential mechanism to choose an outcome $o \in \mathcal{O}$, with quality score $F$. For privacy parameter $\epsilon$, this would give a distribution $\mathcal{M}_\epsilon$ defined to be $\Pr[\mathcal{M}_\epsilon = o] \propto \exp\left(\frac{\epsilon F(o)}{2n}\right)$. Moreover, this mechanism has good social welfare properties: with probability $1-\beta$, it selects some $o$ such that: $F(o) \geq F(o^*) - \frac{2}{\epsilon n}\left(\ln \frac{|\mathcal{O}|}{\beta}\right)$. But as we saw, differential privacy only gives $\epsilon$-approximate truthfulness.

However, \cite{HK12} show that $\mathcal{M}_\epsilon$ is the solution to the following exact optimization problem:
$$\mathcal{M}_\epsilon = \mathrm{arg}\max_{\mathcal{D} \in \Delta\mathcal{O}} \mathbb{E}_{o \sim \mathcal{D}}[F(o)] + \frac{2}{\epsilon n}H(\mathcal{D})$$
where $H$ represents the \emph{Shannon Entropy} of the distribution $\mathcal{D}$. In other words, the exponential mechanism is the distribution which exactly maximizes the expected social welfare, \emph{plus} the entropy of the distribution weighted by $2/(\epsilon n)$. This result implies that the exponential mechanism is \emph{maximal in distributional range}, and hence can be paired with payments to make it exactly truthful.%
\footnote{One way to see this is to view the exponential mechanism as exactly maximizing the social welfare in an augmented setting in with an additional player who cares only about entropy. The payments which make this mechanism truthful are the VCG payments for the augmented game.}
Moreover, they show how to charge payments in such a way as to preserve privacy. The upshot is that for any social choice problem, the social welfare can be approximated in a manner that both preserves differential privacy, and is exactly truthful.

\cite{CCKMV13} also give an (almost equivalent) private generalization of the VCG mechanism and show conditions under which it is truthful \emph{even taking into account agent preferences for privacy}. We discuss this in Section \ref{sec:privcost}.

\subsection{The Sensitive Surveyor's Problem}
In this section, we consider the problem of a data analyst who wishes to conduct a study using the private data of a collection of individuals. However, he must \emph{convince} these individuals to hand over their data! Individuals experience costs for privacy loss. The data analyst can mitigate these costs by guaranteeing differential privacy and compensating them for their loss, while trying to get a representative sample of data. We here closely follow a survey of \cite{Roth12}.

Consider the following stylized problem of the sensitive surveyor Alice. She is tasked with conducting a survey of a set of $n$ individuals $N$, to determine what proportion of the individuals $i \in N$ satisfy some property $P(i)$. Her ultimate goal is to discover the true value of this statistic, $s = \frac{1}{n} |\{i \in N : P(i)\}|$, but if that is not possible, she will be satisfied with some estimate $\hat{s}$ such that the error, $|\hat{s} - s|$, is minimized. We will adopt a notion of accuracy based on large deviation bounds, and say that a surveying mechanism is $\alpha$-accurate if $\Pr[|\hat{s}-s| \geq \alpha] \leq \frac{1}{3}$.  The inevitable catch is that individuals value their privacy and will not participate in the survey for free. Individuals experience some \emph{cost} as a function of their loss in privacy when they interact with Alice, and must be compensated for this loss. To make matters worse, these individuals are rational (i.e. selfish) agents, and are apt to misreport their costs to Alice if doing so will result in a financial gain. This places Alice's problem squarely in the domain of mechanism design, and requires Alice to develop a scheme for trading off statistical accuracy with cost, all while managing the incentives of the individuals.

As an aside, this stylized problem broadly relevant to any organization that makes use of collections of potentially sensitive data. This includes, for example, the use of search logs to provide search query completion and the use of browsing history to improve search engine ranking, the use of social network data to select display ads and to recommend new links, and the myriad other data-driven services now available on the web. In all of these cases, value is being derived from the statistical properties of a collection of sensitive data in exchange for some payment.%
\footnote{The payment need not be explicit and/ or dollar denominated--- e.g. it may be the use of a ``free'' service.}.

Collecting data in exchange for some fixed price could lead to a biased estimate of population statistics, because such a scheme will result in collecting data only from those individuals who value their privacy less than the price being offered. To obtain an accurate estimate of the statistic, it is therefore natural to consider buying private data using an auction, which was recently considered in \cite{GR11}. There are two obvious obstacles which one must confront when conducting an auction for private data, and an additional obstacle which is less obvious but more insidious. The first obstacle is that one must have a quantitative formalization of ``privacy'' which can be used to measure agents' costs under various operations on their data. Here, differential privacy provides an obvious tool. For small values of $\epsilon$, because $\exp(\epsilon) \approx (1+\epsilon)$, it is natural to model agents as having some \emph{linear} cost for participating in a private study. We here imagine that each agent $i$ has an unknown value for privacy $v_i$, and experiences a cost $c_i(\epsilon)=\epsilon v_i$ when his private data is used in an $\epsilon$-differentially private manner.%
\footnote{As we will discuss later, this assumption can be problematic.}
The second obstacle is that our objective is to trade off with \emph{statistical accuracy}, and the latter is not well-studied objective in mechanism design.

The final, more insidious obstacle, is that an individual's cost for privacy loss may be highly correlated with his private data itself! Suppose we only know Bob has a high value for privacy of his AIDS status, and do not explicitly know, this is disclosive because Bob's AIDS status is likely correlated with his value for privacy. More to the point, suppose that in the first step of a survey of AIDS prevalence, we ask each individual to report their value for privacy, with the intention of then running an auction to choose which individuals to buy data from. If agents report truthfully, we may find that the reported values naturally form two clusters: low value agents, and high value agents. In this case, we may have learned something about the population statistic even before collecting any data or making any payments--- and therefore, the agents will have already experienced a cost. As a result, the agents may misreport their value, which could introduce a bias in the survey results. This phenomenon makes direct revelation mechanisms problematic, and distinguishes this problem from classical mechanism design.

\subsubsection{Direct Revelation Mechanisms}
Armed with a means of quantifying an agent $i$'s loss for allowing his data to be used by an $\epsilon$-differentially-private algorithm ($c_i(\epsilon) = \epsilon v_i)$, we are almost ready to describe results for the sensitive surveyor's problem. Recall that a differentially private mechanism is some mapping $M:\mathcal{T}^n\rightarrow \mathcal{O}$, for a general type space $\mathcal{T}$. It remains to define what exactly the type space $\mathcal{T}$ is. We will consider two models. In both models, we will associate with each individual a bit $b_i \in \{0,1\}$ which represents whether they satisfy the sensitive predicate $P(i)$, as well as a value for privacy $v_i \in \mathbb{R}^+$.
\begin{enumerate}
\item In the \emph{insensitive value model}, we calculate the $\epsilon$ parameter of the private mechanism by letting the type space be $\mathcal{T} = \{0,1\}$: i.e. we measure privacy cost only with respect to how the mechanism treats the sensitive bit $b_i$, and ignore how it treats the reported values for privacy, $v_i$.%
\footnote{That is, the part of the mapping dealing with reported values  need not be differentially private.}
\item In the \emph{sensitive value model}, we calculate the $\epsilon$ parameter of the private mechanism by letting the type space be $\mathcal{T}= (\{0,1\} \times \mathbb{R}^+)$: i.e. we measure privacy with respect to how it treats the pair $(b_i, v_i)$ for each individual.
\end{enumerate}
Intuitively, the insensitive value model treats individuals as ignoring the potential privacy loss due to correlations between their values for privacy and their private bits, whereas the sensitive value model treats individuals as assuming these correlations are worst-case, i.e., their values $v_i$ are just as disclosive as their private bits $b_i$. \cite{GR11} show that in the insensitive value model, one can derive approximately optimal direct revelation mechanisms that achieve high accuracy and low cost. By contrast, in the \emph{sensitive value model}, no individually rational direct revelation mechanism can achieve any non-trivial accuracy.

Note that here we are considering a setting in which private data and costs are adversarially chosen. If we are willing to assume a known prior on agent costs (but still assume adversarially chosen private bits $b_i$), then it is possible to improve on the results of \cite{GR11}, and derive Bayesian optimal mechanisms for the sensitive survey problem as is done in \cite{RS12}.
\subsubsection{Take it Or Leave it Mechanisms}
Given the impossibility result of \cite{GR11} for the sensitive value model, the immediate question is what is possible in this setting. Two methods have been recently proposed, which we briefly summarize here. Both approaches abandon direct revelation mechanisms in favor of mechanisms which offer individuals take-it-or-leave-it offers, but both also require subtle changes in how individuals' privacy preferences are modeled. Readers are directed to \cite{FL12,LR12} for more details.
\subsubsection*{Circumventing Impossibility with a Sensitive Surveyor}
Suppose an individual is made a take it or leave it offer: ``If you let us use your bit $b_i$ in an $\epsilon$-differentially private manner, I will give you \$10.'' If values are correlated with private data, an agent's response might reveal something about his bit $b_i$ beyond that which is revealed through the differentially private computation. To model such correlations, \cite{FL12} assume that each individual's value $v_i$ is drawn independently from one of two priors: $v_i \sim F_x$ if $b_i = x$ for $x = 0,1$, known to Alice, the surveyor. Under this assumption, Fleischer and Lyu elegantly construct a take-it-or-leave-it offer which an agent can truthfully decide to accept or reject without revealing \emph{anything} about his private bit! The idea is this: Alice may choose some acceptance probability $q \in [0,1]$. She picks $p_0, p_1$ such that $\Pr_{v \sim F_0}[v \leq p_0/\epsilon] = \Pr_{v \sim F_1}[v \leq p_1/\epsilon] = q$. Alice can then offer the following take-it-or-leave-it offer to each agent: ``If you accept the offer and your (verifiable) bit is $x$, I will pay you $p_x$ dollars, for $x= 1,2$.'' The beauty of this solution is that no matter what private bit the agent has, he will accept the offer with probability $q$ (where the probability is over the corresponding prior) and reject the offer with probability $1-q$. Therefore, \emph{nothing} can be learned about his private bit from his participation decision, and so he has no incentive not to respond to the offer truthfully. Using this idea, \cite{FL12} develop approximately optimal mechanisms that can be used if the priors $F_0$ and $F_1$ are known.
\subsubsection*{Circumventing Impossibility with an Insensitive Surveyor}
What if agent costs are determined adversarially, and there are no known priors? \cite{LR12} give an alternative solution for this case. To circumvent the impossibility result of \cite{GR11}, Alice has one additional power: the ability to accost random members of the population on the street, and present them with a take-it-or-leave-it offer. Once individuals are presented with an offer, they are free to accept or refuse as they see fit. However they do not have the option to \emph{not participate}. If they reject the offer, or even just walk away, this is observed by Alice. This can be seen as a weakening of the standard individual rationality condition. Because costs may be correlated with private data, merely by rejecting an offer or walking away, Alice may learn something about the surveyed individual. If the individual rejects the offer, he receives no payment and yet still experiences some cost! This ends up giving a semi-truthfulness guarantee. If Alice makes an offer of $p$ dollars in exchange for $\epsilon$-differential privacy, a rational agent will accept whenever $p \geq \epsilon v_i$. However, rational agents may accept offers that are below their cost--- because they will still experience some cost by walking away. But these deviations away from ``truthfulness'' are in only one direction, and only help Alice, whose aim it is to compute an accurate population statistic, and does not necessarily care about protecting privacy for its own sake. \cite{LR12} thus obtain non-trivial accuracy in the sensitive value model by making Alice insensitive to the privacy concerns of the agents she surveys, by making offers that they can refuse, but can't avoid.
\subsection{Better Measures for the Cost of Privacy}
\label{sec:privcost}
In the previous section, we took the naive modeling assumption that the cost experienced by participation in an $\epsilon$-differentially private mechanism $M$ was $c_i(o, \mathcal{M},t) = \epsilon v_i$ for some numeric value $v_i$. This measure is problematic for several reasons. First, as pointed out by \cite{NOS12}, although differential privacy promises that any agent's loss in utility is \emph{upper bounded} by a quantity that is (approximately) linear in $\epsilon$, there is no reason to believe that agents' costs are \emph{lower bounded} by such a quantity. That is, while taking $c_i(o, \mathcal{M},t) \leq \epsilon v_i$ is well motivated, there is little support for making the inequality an equality. Second, as discussed in \cite{LR12}, \emph{any} privacy measure which is a deterministic function only of $\epsilon$ (not just a linear function) leads to problematic behavioral predictions.

So how else might we model $c_i$?  One natural measure, proposed by \cite{Xiao13}, is the \emph{mutual information} between the reported type of agent $i$, and the outcome of the mechanism.\footnote{The seminal work of \cite{Xiao13} was the first to explore mechanism design with agents who have costs for privacy loss. It proposes several measures of privacy cost, but mutual information was the first, and helped drive subsequent work.} For this to be well defined, we must be in a world where each agent's type $t_i$ is drawn from a known prior, $t_i \sim \mathcal{T}$.  Each agent's strategy is a mapping $\sigma_i : \mathcal{T}\rightarrow \mathcal{T}$, determining what type he reports, given his true type. We could then define
$$c_i(o, \mathcal{M},\sigma) = I(\mathcal{T}; \mathcal{M}(t_{-i}, \sigma(\mathcal{T})),$$
where $I$ is the mutual information between the random variable $\mathcal{T}$ representing the prior on agent $i$'s type, and $\mathcal{M}(t_{-i}, \sigma(\mathcal{T}))$, the random variable representing the outcome of the mechanism, given agent $i$'s strategy.

This measure has significant appeal, because it represents how ``related'' the output of the mechanism is to the true type of agent $i$. However, in addition to requiring a prior over agent types, \cite{NOS12} observe an interesting paradox that results from this measure of privacy loss. Consider a world in which there are two kinds of sandwich breads: Rye (R), and Wheat (W). Moreover, in this world, sandwich preferences are highly embarrassing and held private. The prior on types $\mathcal{T}$ is uniform over R and W, and the mechanism $\mathcal{M}$ simply gives agent $i$ a sandwich of the type that he purports to prefer. Now consider two possible strategies, $\sigma_{\textrm{truthful}}$ and $\sigma_{\textrm{random}}$. $\sigma_{\textrm{truthful}}$ corresponds to truthfully reporting sandwich preferences (and subsequently leads to eating the preferred sandwich type), while $\sigma_{\textrm{random}}$ randomly reports independent of true type (and results in the preferred sandwich only half the time).  The cost of using the random strategy is $I(\mathcal{T}; \mathcal{M}(t_{-i}, \sigma_{\textrm{random}}(\mathcal{T})) = 0$, since the output is independent of agent $i$'s type. On the other hand, the cost of truthfully reporting is $I(\mathcal{T}; \mathcal{M}(t_{-i}, \sigma_{\textrm{truthful}}(\mathcal{T})) = 1$, since the sandwich outcome is now the identity function on agent $i$'s type. However, from the perspective of any outside observer, the two strategies are indistinguishable! In both cases, agent $i$ receives a uniformly random sandwich. Why then should anyone choose the random strategy? So long as an adversary \emph{believes} they are choosing randomly, they should choose the honest strategy.

\cite{CCKMV13} propose a different approach, not needing a prior on agent types. They propose the following  cost function:\footnote{In fact, the upper bound they propose is more general, replacing the $\ln(\cdot)$ used here with any function $F_i$ satisfying certain conditions.}
$$\left|c_i(o, \mathcal{M},t)\right| = \ln\left(\max_{t_i, t_i' \in \mathcal{T}}\frac{\Pr[\mathcal{M}(t_i, t_{-i}) = o]}{\Pr[\mathcal{M}(t'_i, t_{-i}) = o]}\right).$$
Note that if $\mathcal{M}$ is $\epsilon$-differentially private, then
$$ \max_{t \in \mathcal{T}^n}\max_{o \in \mathcal{O}}\max_{t_i, t_i' \in \mathcal{T}}\ln\left(\frac{\Pr[\mathcal{M}(t_i, t_{-i}) = o]}{\Pr[\mathcal{M}(t'_i, t_{-i}) = o]}\right) \leq \epsilon.$$
That is, we can view differential privacy as bounding the \emph{worst-case} privacy loss over all possible outcomes, whereas the measure proposed by \cite{CCKMV13} considers only the privacy loss for the outcome $o$ (and type vector $t$) actually realized. Thus, for any differentially private mechanism $\mathcal{M}$,  $\left|c_i(o, \mathcal{M},t)\right| \leq \epsilon$ for all $o, t$, but it will be important that the cost can vary by outcome.

\cite{CCKMV13} then consider the following allocation rule for maximizing social welfare $F(o) = \sum_{i=1}^n u_i(o).$%
\footnote{This allocation rule is extremely similar to, and indeed can be modified to be identical to the exponential mechanism.}
We discuss the case when $|\mathcal{O}| = 2$ (which does not require payments), but the authors analyze the general case (with payments), which privately implements the VCG mechanism for any social choice problem.
\begin{enumerate}
\item For each outcome $o \in \mathcal{O}$, choose a random number $r_o$ from the distribution $\Pr[r_o = x] \propto \exp(-\epsilon |x|)$.
\item Output $o^* = \mathrm{arg}\max_{o \in \mathcal{O}}(F(o) + r_o)$
\end{enumerate}
\cite{CCKMV13} show that the above mechanism is $\epsilon$-differentially private, and that it is truthful for privacy aware agents, so long as for each agent $i$, and for the two outcomes $o, o' \in \mathcal{O}$,  $|\mu_i(o) - \mu_i(o')| > 2\epsilon$. Note that this will be true for small enough $\epsilon$ so long as agent utilities for outcomes are distinct. The analysis is elegant, and proceeds by considering an arbitrary fixed realization of the random variables $r_o$, and an arbitrary deviation $t'_i$ from truthful reporting for the $i$'th agent. There are two cases: In the first case, the deviation does not change the outcome $o$ of the mechanism. In this case, \emph{neither} the agent's utility for the outcome $\mu_i$, nor his cost for privacy loss $c_i$ change at all, and so the agent does not benefit from deviating. In the second case, if the outcome changes from $o$ to $o'$ when agent $i$ deviates, it must be that $\mu_i(o') < \mu_i(o) - 2\epsilon$. By differential privacy, however, $|c_i(o, \mathcal{M},t) - c_i(o', \mathcal{M},t)| \leq 2\epsilon$, and so the change in privacy cost cannot be enough to make it beneficial.

Finally, \cite{NOS12} take the most conservative approach to modeling costs for privacy. Given an $\epsilon$-differentially private mechanism $\mathcal{M}$, they assume that
$$c_i(o, \mathcal{M},t) \leq \epsilon v_i,$$
for some number $v_i$. This is similar to the linear cost functions of \cite{GR11} that we considered earlier, but crucially, \cite{NOS12} assume only an upper bound. This assumption is satisfied by all of the other models for privacy cost that we have considered thus far. They show that many mechanisms that combine a differentially private algorithm with a punishing mechanism that has the ability to restrict user choices, like those from \cite{NST12} that we considered in Section \ref{sec:restrict}, maintain their truthfulness properties in the presence of agents with preferences for privacy, so long as the values $v_i$ are bounded. Moreover, they go on to show that for a great many distributions from which the values $v_i$ might be drawn (even with \emph{unbounded support}), it is still possible to make truthful reporting a dominant strategy for \emph{almost} all agents, which is often sufficient to get strong welfare guarantees.

\section{Conclusions and Open Questions}
The science of privacy, and its burgeoning connections to game theory and mechanism design are still in their very early stages. As such, there remain more questions than answers at this intersection, but there is already more than enough evidence that the area is rich, and that the answers will be fascinating. We here suggest just a couple of the high-level questions that deserve to be better understood:
\begin{enumerate}
\item Why do people care about privacy? In Section \ref{sec:value} we began to see some answers to this question: people might care about privacy, because of some \emph{repeated interaction}, in which the revelation of their type at one stage of the interaction might cause them quantifiable harm at some later stage of the game. This literature has already provided us with some insights, but only at a very coarse-grained level. The recent literature on differential privacy lets us discuss privacy as a quantitative, rather than just qualitative property. Can we formulate a rich model in which we can directly derive agents' values for (differential) privacy in various settings? This study will surely help us understand how we should in general model agent costs for privacy when trying to design truthful mechanisms for agents who explicitly experience costs for privacy in their utility functions, as we considered in Section \ref{sec:aware}.
\item How should we model privacy costs that are not captured by information theoretic measures like differential privacy, but nevertheless seem to have real economic consequences? In particular, how can we model the fact that \emph{search costs} appear to factor into peoples' perceptions of privacy? For example, there has been a recent uproar about Facebook's new Graph Search tool.\footnote{See e.g. \url{http://www.slate.com/blogs/future_tense/2013/01/23/actual_facebook_graph_searches_tom_scott_s_tumblr_a_privacy_wake_up_call.html}} From the perspective of differential privacy, the addition of this feature, which merely eases the search through information that was already \emph{in principle} publicly available, should have had no incremental privacy cost. Nevertheless, it plainly does. Is there a clean economic model that captures this?
\item In settings in which agents care about privacy, to what extent does dominant strategy truthfulness remain the right solution concept? Depending on the privacy model chosen, it is no longer clear that the revelation principle holds in such settings, because agents have preferences not only over the outcome chosen by the mechanism, but also over the mechanism which chooses the outcome itself. Perhaps in such settings, \emph{better} outcomes can be implemented in Nash equilibrium than can be implemented in dominant strategies?
\item Differential privacy is clearly a powerful tool for reasoning about noise and stability in mechanism design. Already in Section \ref{sec:tool}, we saw several examples of results easily derived via differential privacy, which were are not known how to accomplish in any other way. It seems like a particularly promising tool for reasoning about asymptotic truthfulness, which is a compelling second-best solution concept when exactly dominant strategy truthful mechanisms are not known. Can we use differential privacy to map out the power of asymptotically truthful mechanisms? How much more powerful is this class as compared to the set of exactly truthful mechanisms?
\item In several settings of applied interest, a regulator may wish to reveal summary information about the industry she regulates to guide policy and alleviate information asymmetries. In doing this, she must balance the privacy of individual participants, which arise from competitive concerns, trade secrets etc. Can differential privacy and its variants assist in the design of such information release? How should policy be designed given that the summary used to guide it is privacy-preserving and therefore necessarily coarse? \cite{flood2013supervisory} propose and study these concerns in the context of the financial industry, and provide an excellent overview of the trade-offs involved.
\end{enumerate}

\bibliographystyle{alpha}
\bibliography{refs}

\end{document}